# Thermodynamics of Benford's First Digit Law

By Don S. Lemons

Bethel College

North Newton, Kansas

16 June 2015

Abstract

Iafrate, Miller, and Strauch ["Equipartition and a Distribution for Numbers: A Statistical Model for Benford's Law," forthcoming in Physical Review E] construct and test a statistical model of the partitioning of a conserved quantity. One consequence of their model is Benford's law according to which the frequency of appearance of the first digit $d$ of numbers taken from a set of data is $\log_{10}(1+1/d)$. This Comment amplifies their work by exploring the thermodynamic consequences of their statistical model.

What distinguishes those data sets that do and those that do not observe Benford's law? Recently Iafrate, Miller, and Strauch[1] have verified an earlier suggestion by the current author,[2] that Benford's law[3] results from the random partitioning of a conserved quantity. This Comment makes explicit the thermodynamics of Benford's law implied by Iafrate et. al.'s statistical model. First I present a brief version of their statistical model. Then I develop its thermodynamic consequences.

Iafrate et. al. use the principle of maximum entropy to determine the probability of partitioning a conserved quantity $X$ into pieces of size $i\Delta x$ where $i = 1, 2, \ldots N$. In general, the numbers of pieces of size $i\Delta x$ range from $0$ to $\infty$ for each $i$. Thus I denote the possible numbers of pieces of size $i\Delta x$ with $n_{i,j} = j$ for $j = 0, 1, \ldots \infty$ and each $i$. All such partitions must conserve the quantity $X$ where

$$X = \sum_{i=1}^{N} x_i \sum_{j=0}^{\infty} p_i(n_{i,j}) n_{i,j} . \tag{1}$$

Upon maximizing the entropy

$$S = \sum_{i=1}^{N} S_i \tag{2}$$

where

$$S_i = -\sum_{j=0}^{j=\infty} p_i(n_{i,j}) \ln p_i(n_{i,j}) \tag{3}$$

subject to normalized probabilities $p_i(n_{i,j})$ and to constraint (1), Iafrate et. al. find that

$$p_i(n_{i,j}) = e^{-\beta i n_{i,j}}\left(1 - e^{-\beta i}\right) \tag{4}$$

where $n_{i,j} = j$ and $\beta$ is a Lagrange multiplier whose value is determined by imposing constraint (1). Therefore the probability of a particular partition, described by a particular set of numbers $n_{i,j}$ for which $\sum_{i=1}^{N} x_i n_{i,j} = X$, is given by

$$\prod_{i=1}^{N} p_i(n_{i,j}) = e^{-\beta X/\Delta x}\left(e^{\beta} - 1\right)\left(e^{2\beta} - 1\right)\ldots\left(e^{N\beta} - 1\right). \tag{5}$$

Thus each partition is equiprobable with a probability given by the right hand side of (5). Subsequently, Iafrate et. al. compute the average number of pieces of size $i\Delta x$,

$$\langle n_i \rangle = \sum_{j=0}^{\infty} p_i(n_{i,j}) n_{i,j} , \tag{6}$$

from (4) and find that

$$\langle n_i \rangle = \frac{1}{e^{\beta i} - 1} . \tag{7}$$

Iafrate et. al. call "the usual case in real-world data," in which the conserved quantity $X$ is composed of many more parts $X/\Delta x$ than piece sizes $N$ the "equipartition limit." In this regime $\beta N \ll 1$ and, given constraint (1), $\beta = N\Delta x/X$, and so $X/\Delta x \gg N^2$. Then (7) reduces to

$$\langle n_i \rangle = \frac{X}{iN\Delta x} . \tag{8}$$

The dependence $\langle n_i \rangle \propto 1/i$ is called "Benford" since its normalized integration over an arbitrary decade of closely spaced piece sizes,

$$\frac{\int_{d10^p}^{(d+1)10^{p+1}} dx/x}{\int_{10^p}^{10^p} dx/x} = \log_{10}\left(1+\frac{1}{d}\right), \qquad (9)$$

yields the Benford frequency of first digit $d$.

In Iafrate et. al.'s statistical analysis the numbers $n_{i,j} = j$ with $j = 0,1,2,\ldots\infty$ are the values assumed by $N$ random numbers defined by probabilities (4). In contrast, it is their mean values $\langle n_i \rangle$ that are the macroscopic variables that determine the state of a thermodynamic system. In conformity with thermodynamic usage, $n_i$ here and below replaces what before has been denoted $\langle n_i \rangle$.

Consider then that each set of $n_i$ pieces of size $x_i$ $[= i\Delta x]$ composes a subsystem with its own entropy $S_i$ and portion $n_i i\Delta x$ of the conserved quantity $X$. According to (1), in thermodynamic notation

$$X = \sum_{i=1}^{N} in_i \Delta x, \qquad (10)$$

and (2) the subsystem quantities $n_i i\Delta x$ and $S_i$ are additive over the system. While additive, the entropy of the thermodynamic system is not an extensive function of extensive variables because the system's $N$ parts are not simply small versions of the whole. The entropy of each subsystem $i$ is determined by using probabilities (4) to complete the sum in (3). In this way we find that

$$S_i = \ln n_i + n_i \ln n_i - n_i \ln(n_i - 1) \qquad (11)$$

describes the entropy of subsystem *i*. The derivative of $S_i$ with respect to its additive variable $in_i \Delta x$ must be an intensive variable, called $\beta$, whose equality among the different subsystem is required for their equilibrium. Thus, the equations of state

$$\frac{1}{i\Delta x}\left(\frac{\partial S_i}{\partial n_i}\right) = \beta \qquad (12)$$

reproduce Iafrate et. al.'s result (7). [Maximizing the system entropy $S$ subject to constraint (10) also produces (7).]

The equipartition limit requires that $n_i \gg 1$ for all *i*. In this regime, the subsystem entropy (11) reduces to

$$S_i = \ln(en_i) \qquad (13)$$

and the entropy of the composite system (2) to

$$S = \ln(e^N n_1 n_2 \cdots n_N) \qquad (14)$$

The equation of state (12) applied to subsystem entropy (13) produces the Benford distribution (8). Note that the argument of the logarithm of (14) is formally a system multiplicity $\Omega$. The Benford distribution (8) allows us to express this multiplicity as

$$\Omega = \left(\frac{Xe}{N\Delta x}\right)^N \frac{1}{N!} \ . \qquad (15)$$

That this multiplicity $\Omega$ is, in fact, identical to the number of system partitions is clear from the following. Since the probability of any one partition (5) is the same as any other, the inverse of this probability must be the total number of partitions $\Omega_p$ consistent with $X/\Delta x$ and $N$. Therefore, given (5), in general,

$$\Omega_p = \frac{e^{\beta X/\Delta x}}{\left(e^{\beta}-1\right)\left(e^{2\beta}-1\right)\ldots\left(e^{N\beta}-1\right)} \tag{16}$$

where $\beta$ is determined by (10). In the equipartition regime, where $N\beta \ll 1$ and $\beta = N\Delta x/X$, the right hand side of (16) reproduces the right hand side of (15) and so, at least in this limit, $\Omega_p = \Omega$.

Iafrate et. al.'s largest simulation is one for which $X/\Delta x = 800$ and $N = 10$. Placing these data into (15), as transformed by Stirling's approximation into

$$\Omega_p = \left(\frac{Xe^2}{N^2 \Delta x}\right)^N \frac{1}{\sqrt{2\pi N}} \;, \tag{17}$$

produces a multiplicity, $600 \cdot 10^{14}$, about 600 times larger than the number of partitions reported -- assuming "over 100 trillion" means "only a little over 100 trillion." This mismatch may arise because the condition $X/\Delta x \gg N^2$ required by (17) is not sufficiently realized by $800 \gg (10)^2$. If so, the number of partitions, especially those involving large pieces of size $i\Delta x$ where $i \approx N$, will be depressed below those predicted by (17).

Given the non-intuitive nature of the system entropy (14) in the equipartition limit, it seems unlikely that the thermodynamics of Benford's law, as outlined here, can be motivated on any other ground than that provided by the elegant statistical analysis of Iafrate, Miller, and Strauch.

---

[1] Joseph B. Iafrate, Steven J. Miller, and Frederick W. Strauch, "Equipartitions and a Distribution for Numbers: A Statistical Model for Benford's Law," arXiv:1503.08259. (forthcoming in Physical Review E)

---

[2] Don. S. Lemons, "On the Numbers of Things and the Distribution of First Digits," American Journal of Physics 54(9), pp. 816-817. (1986).

[3] Simon Newcomb first discovered "Benford's law." See his "Note on the frequency of use of the different digits in natural numbers," American Journal of Mathematics 4(1), 39-40, (1881). Frank Benford's article, "The law of anomalous numbers." Proceedings of the American Philosophical Society, Vol. 78, No. 4, pp. 551-572, (March 31, 1938), subsequently popularized the law.